\documentclass{PoS}
\usepackage{graphicx}% Include figure files
\usepackage{amsmath}
\usepackage{float}
\usepackage{perpage}
\MakeSorted{figure}
\MakeSorted{table}
\usepackage{morefloats}

\RequirePackage{xspace}
\def\epem {\ensuremath{e^+e^-}\xspace}
\def\gg {\ensuremath{\gamma \gamma}\xspace}
\def\ge {\ensuremath{\gamma e}\xspace}

\def\invfb {\ensuremath{\mathrm{fb}^{-1}}\xspace}
\def\invab {\ensuremath{\mathrm{ab}^{-1}}\xspace}
\def\ggr {\ensuremath{{\rm  \gamma\gamma}}\xspace}

\def\mumur {\ensuremath{{\rm \mu\mu}}\xspace}
\def\tautau {\ensuremath{{\rm   \tau\tau}}\xspace}

\def\t-t {\ensuremath{{\rm   tt}}\xspace}

\def\inv {\ensuremath{{\rm   inv}}\xspace}

\def\G {\ensuremath{{\rm   \Gamma}}\xspace}
\def\dG {\ensuremath{{\rm   \Delta \Gamma}}\xspace}
\newcommand{\mev}{\ensuremath{\mathrm{\,Me\kern -0.1em V}}\xspace}
\newcommand{\mevcc}{\ensuremath{{\mathrm{\,Me\kern -0.1em V\!/}c^2}}\xspace}
\newcommand{\gev}{\ensuremath{\mathrm{\,Ge\kern -0.1em V}}\xspace}
\newcommand{\gevcc}{\ensuremath{{\mathrm{\,Ge\kern -0.1em V\!/}c^2}}\xspace}
\newcommand{\tev}{\ensuremath{\mathrm{\,Te\kern -0.1em V}}\xspace}
\newcommand{\tevcc}{\ensuremath{{\mathrm{\,Te\kern -0.1em V\!/}c^2}}\xspace}
\newcommand{\ev}{\ensuremath{\mathrm{\,e\kern -0.1em V}}\xspace}

\def\mum  {\ensuremath{{\,\mu\rm m}}\xspace}%% mu meter

\def\cms  {\ensuremath{{\rm \,cm}^{-2} {\rm s}^{-1}}\xspace}

\title{\boldmath Higgs factories}

\ShortTitle{Higgs factories}

\author{
        \speaker{V.~I.~Telnov}\\
        Budker Institute of Nuclear Physics, SB RAS, 630090, Novosibirsk, Russia\\
        Novosibirsk State University, 630090, Novosibirsk, Russia\\
        E-mail: \email{telnov@inp.nsk.su}
}

\abstract{Over the past two decades, the high energy physics community has been actively discussing and developing a number of post-LHC collider projects; however, none of them have been approved due to high costs and the uncertainty in post-LHC physics scenarios.  There have been great expectations of rich new physics in the 0.1--1 \tevcc mass region: the Higgs boson (one or several), supersymmetry, or perhaps new particles from the dark-matter family. It has been the general consensus that the best machine for the detailed study of new physics to be discovered at the LHC would be an energy-frontier linear \epem collider.  Physicists held their breath waiting for the results from the LHC. In summer 2012, two LHC detectors, ATLAS and CMS, announced the discovery of a Higgs boson with the mass of 126 \gevcc---and (still) nothing else. The absence of new physics in the region below 1 \tevcc has changed the post-LHC collider R\&D priorities and triggered a zoo of project proposals for the precision study of the 126 \gevcc Higgs boson, possibly with further upgrades to higher energies. This paper gives an overview of these projects; it is based largely on the reports presented at the first workshop on Higgs factories held at FNAL a few days prior to the present workshop in Protvino.
}

\FullConference{LHC on the March - IHEP-LHC,\\
        20-22 November 2012\\
        Institute for High Energy Physics, Protvino, Moscow region, Russia }
% PACS,

\begin{document}
\section{Introduction}

\noindent
In the middle of 2012, the ATLAS~\cite{higgsATLAS} and CMS~\cite{higgsCMS} detectors at LHC announced the discovery of a new particle with the mass of about 126 \gevcc, with properties consistent with those predicted for the Standard Model Higgs boson. This discovery, without a doubt, is a huge success for the LHC and for high energy physics as a whole. Physicists around the world had been waiting for many years for the first round of LHC discoveries in order to decide what the next HEP projects should be. At \textit{Snowmass 2001}, a large forum on the future of particle physics, the HEP community was unanimous that the next large HEP project should be a linear collider (LC) with the energy $2E_0$=500--1000 \gev. At present, two linear collider projects remain: ILC ($2E_0$=250--1000 \gev) and CLIC ($2E_0$=350--3000 \gev). The ILC team has already prepared a Technical Design Report and is ready to start construction, while the CLIC team recently issued a Conceptual Design, with a Technical Design to be ready a few years later. Due to high costs, it is clear that no more than one LC can be build---but which one?

Up to now, the LHC has found only the Higgs boson---and nothing else below approximately 1 \tevcc: no supersymmetry, no dark matter particles, not a hint of anything else. It is not excluded that new physics will yet be found at LHC as higher statistics are accumulated and as LHC ramps up to its full design energy of 14 TeV.  This means that the LC decision could be made no earlier than 2018. Two years ago, before the Higgs boson was discovered at the LHC, this was an official ICFA opinion. Now, the situation has changed. The physics motivation for an energy-frontier LC is no longer as strong as before because we know that the energy region below 1 TeV is not nearly as rich as had been expected. Are there any other strategies HEP could follow? For example, let us consider the possibility of building a low-energy facility for the detailed study of the Higgs boson while leaving the energy frontier to the LHC, the high-luminosity HL-LHC, and to some future, even more high-energy $pp$ or muon collider.

At first sight, even in this scenario, the ILC is the top candidate for a Higgs factory, ready to go; if not the ILC---what else? In December 2011, the day the first indication of a $\sim 125 \gevcc$ resonance was announced, A.~Blondel and F.~Zimmermann published an e-print~\cite{Zim} with the proposal of an \epem ring collider in the LHC tunnel, dubbed LEP3, to study the Higgs boson. That e-print triggered a strong renewed interest in \epem ring colliders because for such a low-mass Higgs boson one needs a ring with the energy $2E_0 = 240 \gev$, which is only somewhat higher than the LEP2 peak energy (209 GeV). The luminosity of such a ring collider could be comparable to that at the ILC. Soon thereafter, it became clear that it would be preferable to build a ring collider with a radius several times as large as LEP-2's because: a) for a fixed power, the luminosity is proportional to the ring's radius, b) in the future, one can place in the same tunnel a $\sim 100$ \gev $pp$ collider. This clearly sounds like a serious long-term HEP strategy. Another option is a muon-collider Higgs factory. The technology is not ready yet, but the development of muon colliders is needed in any case for access to the highest energies; therefore, such a strategy makes sense as well. There are also suggestions for a ring-type photon collider Higgs factory (without \epem) based on recirculating linacs, though usually photon colliders are considered as a natural add-on to \epem\ linear colliders.

This conference paper is based on materials of the recent first Higgs factory workshop HF2012, held at Fermilab in November 2012~\cite{Blo2},
where various Higgs factory proposals have been presented and compared:

\begin{itemize}
\item Proton colliders
\begin{itemize}
\item a) LHC, b) HL-LHC, c) HE-LHC, d) SHE-LHC, e) VLHC.
\end{itemize}
\item Linear \epem colliders
\begin{itemize}
\item a) ILC, b) CLIC, c) X-band klystron-based.
\end{itemize}
\item Circular \epem colliders
\begin{itemize}
\item a) LEP3, b) TLEP (Triple-size LEP)~\cite{Blo}, c) SuperTRISTAN-40(80),  d) Fermilab
site-filler, e) CHF-1 and CHF-2 (China), f) VLLC~\cite{Sum}.
\end{itemize}
\item Photon colliders
\begin{itemize}
\item a) ILC-based, b) CLIC-based, c) Recirculating linac-based (SAPPHiRE), d) SLC-type.
\end{itemize}
\item Muon collider
\end{itemize}

Below, we review the physics motivation of each Higgs factory option and then consider key technical features of each project.

\section{Higgs physics}
\noindent
The Higgs boson has been detected at the LHC in multiple channels: $bb$, $\tau\tau$, $\gg$, $WW$, $ZZ$. The measurement of the Higgs mass gave us
the last unknown parameter of the Standard Model (AM). Now, all SM cross sections and branching ratios can be calculated
and compared with the experiment. Any statistically significant deviation would signal the existence of new physics. 
What Higgs properties should be measured and what accuracy is needed?

According to theoretical predictions, new physics appearing at the 1 TeV scale could change the Higgs branching ratios by $\sim 1$--5\%. Detecting such a small discrepancy would require a branching-ratio measurement precision of much better than one percent. The leading branchings of the 126 \gev Higgs boson are shown in Fig.~\ref{h-branch}.  One could also measure the Higgs' branching ratio to $\mu\mu$, which is about 0.022\%; branchings to \epem\ and light quarks are too small to be measured. The total Higgs width is about 4 MeV, it can be measured at \epem and muon colliders (see below). Electron-positron colliders also present a nice possibility to measure the Higgs decay width to invisible states. The Higgs coupling to the top quark can be measured at both the LHC and high-energy \epem colliders. The Higgs self-coupling is very important but difficult to measure; its measurement requires either the high-luminosity LHC or a high-energy linear \epem collider. Let us consider the Higgs physics accessible at each type of colliders.

\subsection{The LHC as a Higgs factory}
\begin{figure}[!tbp]
     \begin{center}
     \vspace*{-0.2cm}
\includegraphics[width=6.5cm] {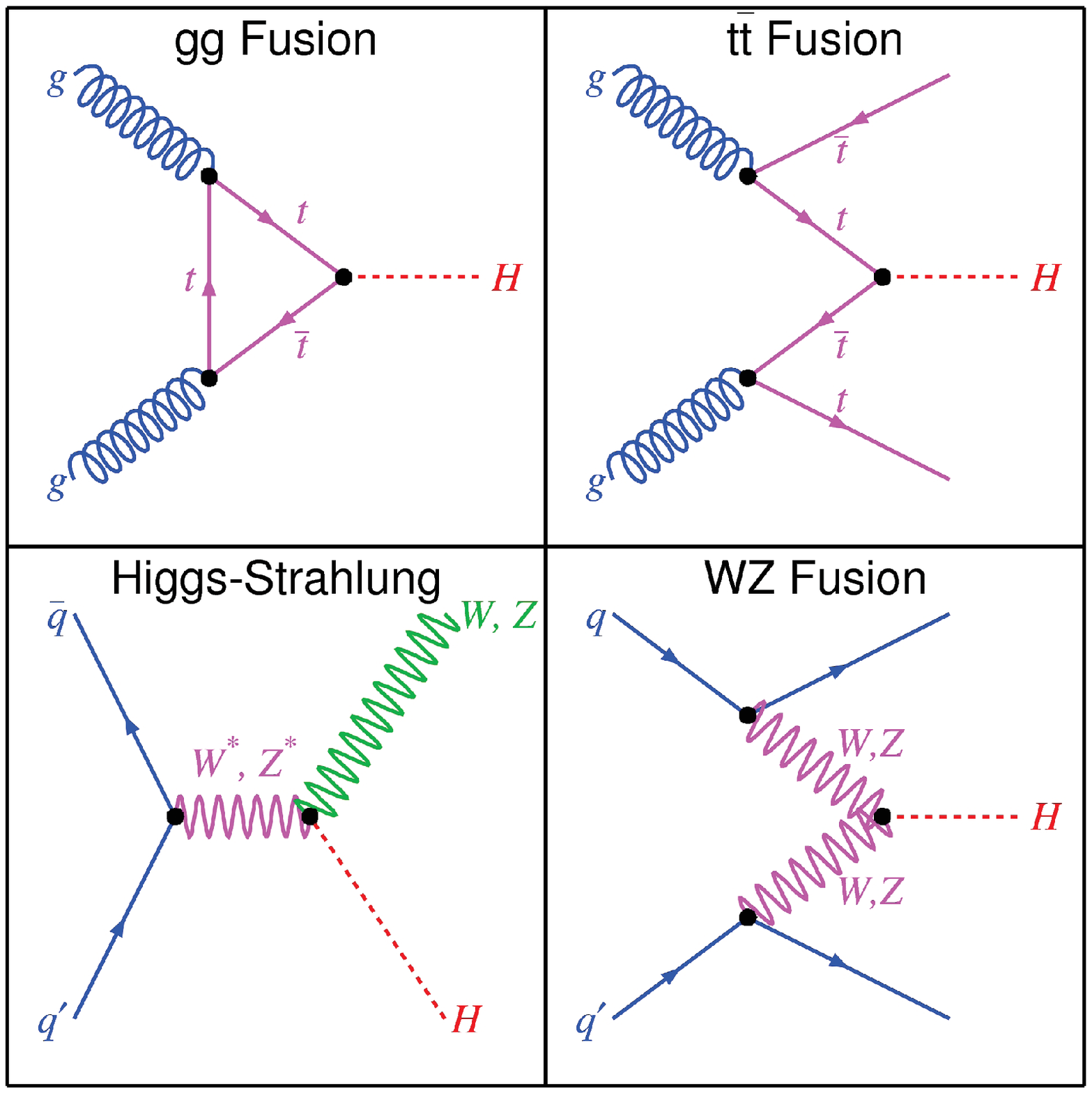}
\includegraphics[width=8.5cm] {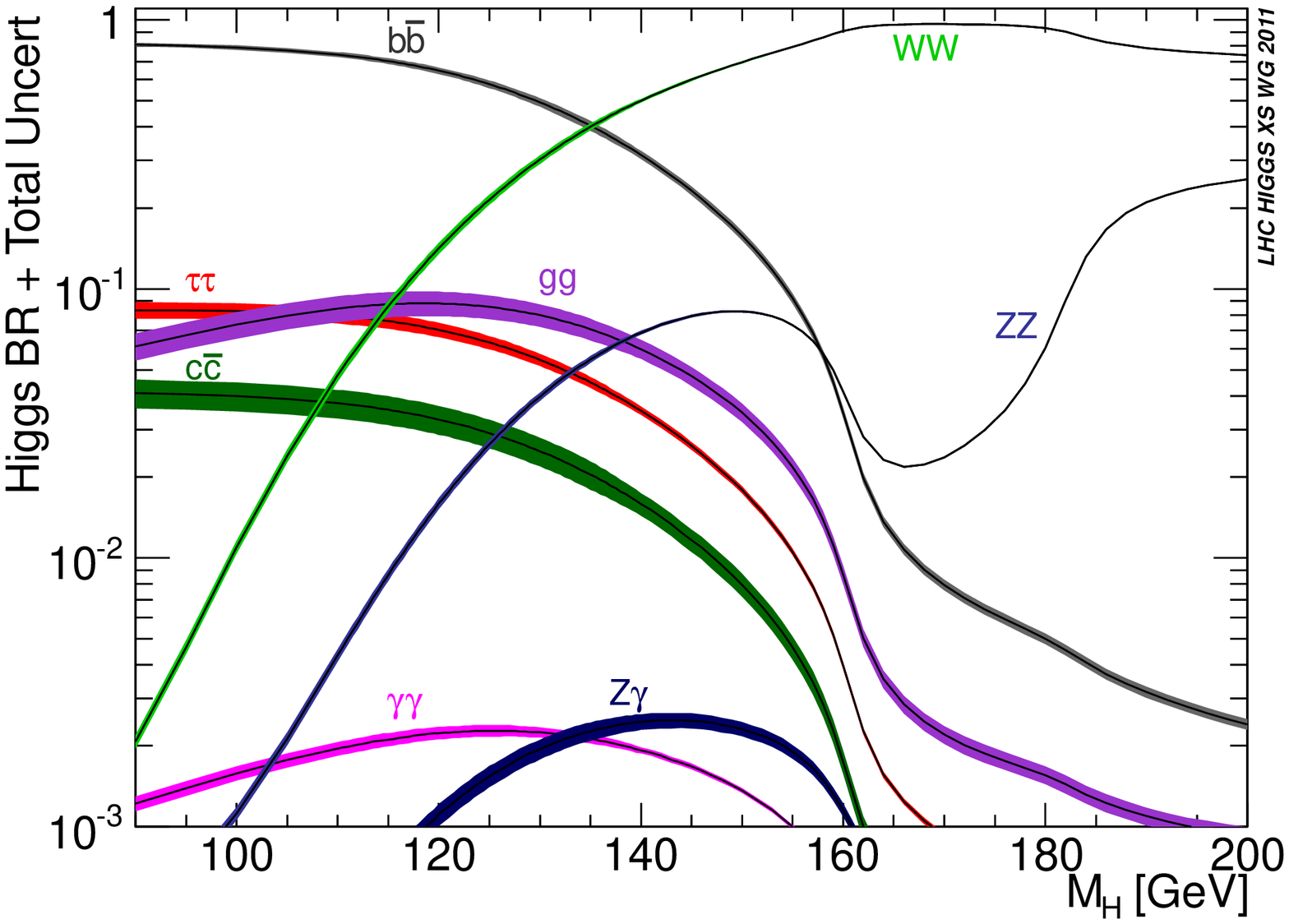}
       \vspace*{-0.7cm}
     \end{center}
     \caption{Left: diagrams for the Higgs boson production in $pp$ collisions; right: the Higgs boson branchings. }
   \vspace*{0.5cm}
   \label{h-branch}
   \end{figure}

\begin{figure}[!tbp]
     \begin{center}
     \vspace*{-0.2cm}
   \includegraphics[width=7.5cm] {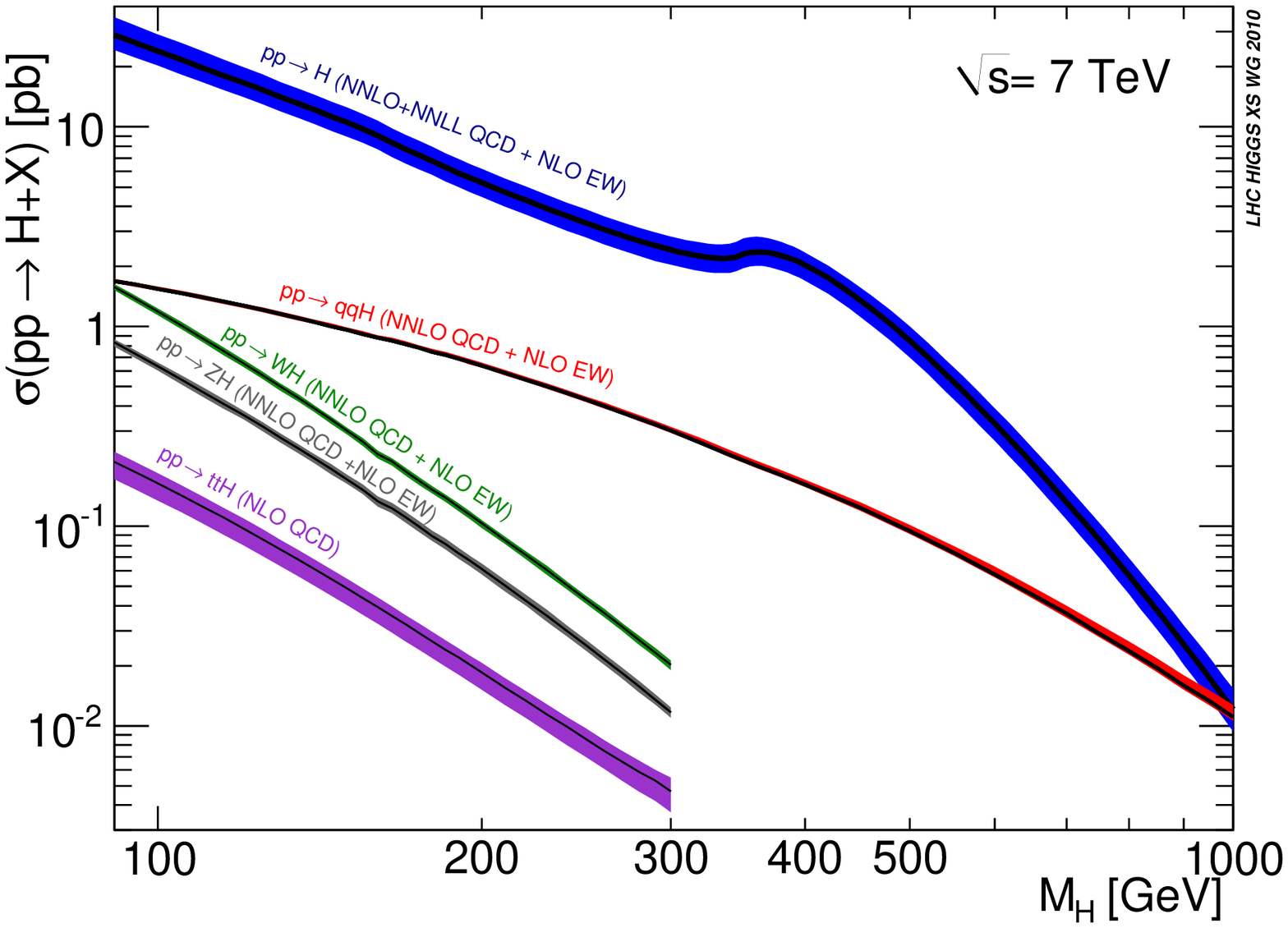} \includegraphics[width=7.5cm] {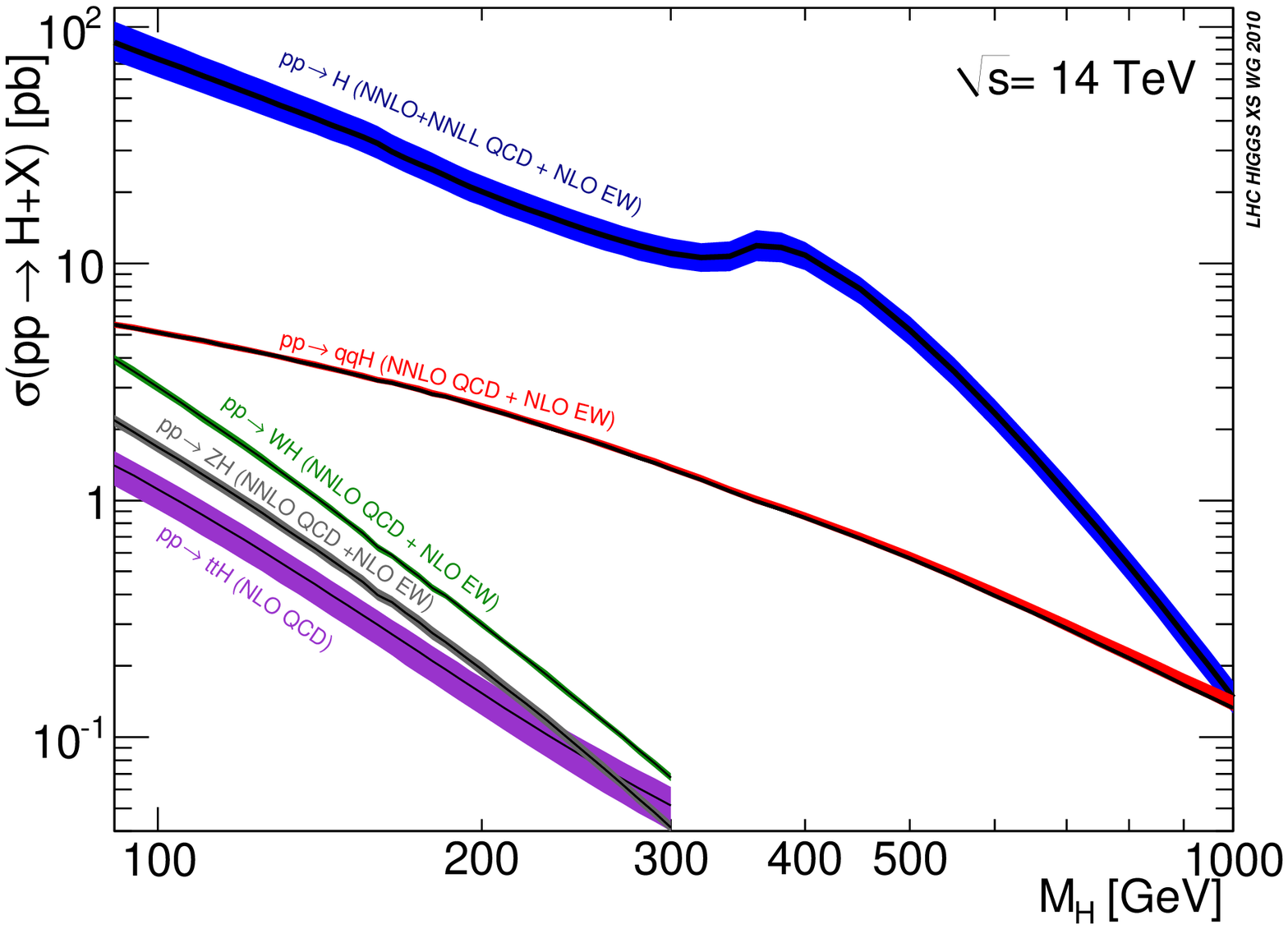}
     \vspace*{-0.7cm}
     \end{center}
     \caption{Cross section for the Higgs boson production in $pp$ collisions.}
   \vspace*{0.5cm}
   \label{h-cross-pp}
   \end{figure}

\noindent
The cross sections for Higgs boson production in $pp$ collisions at c.m.s. energies of 7 and 14 TeV are shown in Fig.\ref{h-cross-pp}. The total cross section at 14 TeV is about 57 pb. The expected integrated luminosity is 300 \invfb at the nominal LHC and 3000 \invfb at the high-luminosity HL-LHC, which corresponds to the production of about 20 million and 200 million Higgs bosons, respectively, many more than at any other Higgs factory. The main problems at the LHC are backgrounds and the uncertainties of the initial state and the production mechanisms. Higgs production mechanisms that apply to $pp$ collisions are shown in Fig.~\ref{h-branch} (left), and their cross sections are shown in Fig.~\ref{h-cross-pp}.
The Higgs boson is produced in gluon fusion, vector-boson fusion, and radiation from top quark, $W$ or $Z$ bosons. For each final state, one can identify the initial state by kinematic selection. This way, one can measure the $Htt$ coupling. The Higgs boson can be detected in all final states enumerated above, with the exception of $c\bar{c}$. The total and invisible Higgs widths cannot be measured in $pp$ collisions.

\subsection{\boldmath{Higgs physics at \epem colliders}}
\begin{figure}[tbp]
     \begin{center}
%     \vspace*{-0.2cm}
\includegraphics[height=6.2cm,clip=true] {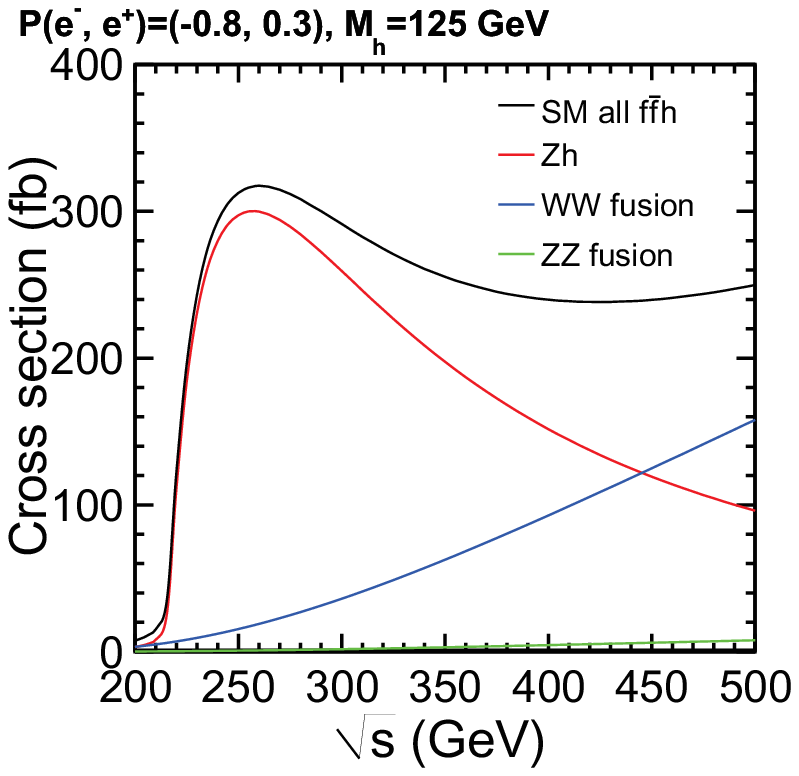} ~~~~ \includegraphics[height=5.9cm,clip=true] {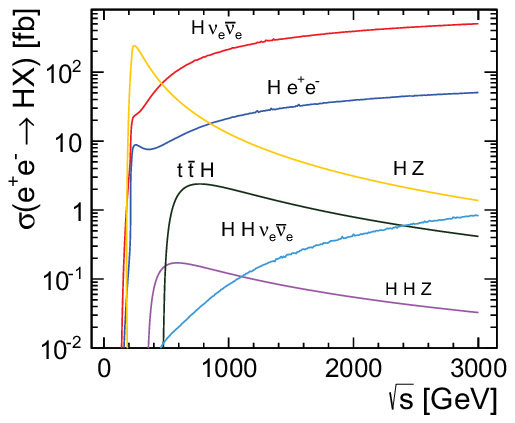} \\
\vspace*{0.2cm}
\includegraphics[width=12cm,clip=true] {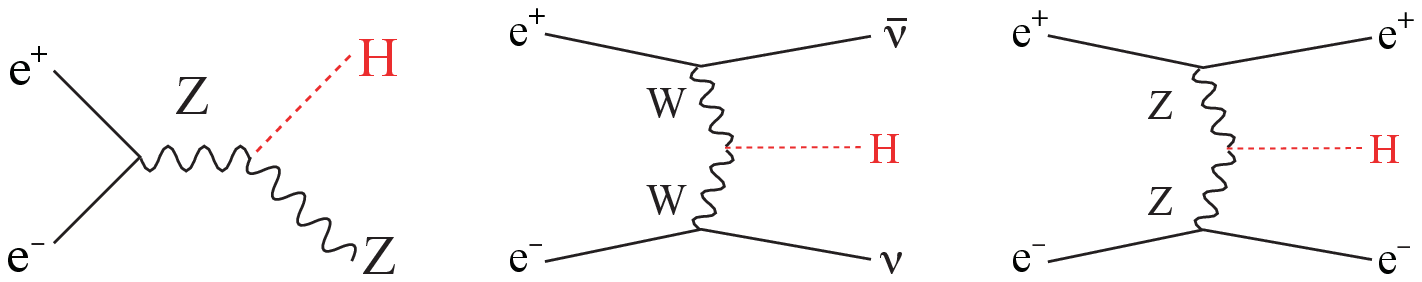}
       \vspace*{-0.3cm}
     \end{center}
     \caption{Cross sections and leading diagrams for the Higgs boson production in \epem collisions.}
   \vspace*{0.5cm}
   \label{h-cross-e+e-}
   \end{figure}

\noindent
Cross sections and the leading diagrams for Higgs boson production in \epem collisions are shown in Fig.~\ref{h-cross-e+e-}. The total cross section at the 240 GeV \epem Higgs factory is about 300 fb (200 fb for unpolarized beams). Typical luminosity of \epem colliders (LC or rings) is about $10^{34}$ \cms, therefore, the total number of produced Higgs bosons per one year ($10^7$ sec) is about 20000-30000, or about 100000 for life of the experiment. Large circular \epem colliders like TLEP (C$=80$ km) or IHEP (C$=70$ km) can have  luminosities several times greater than at linear colliders. In addition, ring colliders can have several interaction points; therefore, the number of produced Higgs bosons could reach 1 million.

   A unique feature of \epem colliders is the reaction $\epem \to ZH$. By detecting the leptonic decays of the $Z$, one can measure directly all branching ratios and see even invisible Higgs boson decays via the recoil mass. The total cross section for this reaction is proportional to $g_{HZZ}^2$, while the cross section with decay $H\to ZZ$ is proportional to $g_{HZZ}^4/\Gamma_H$. Measurements of this cross section and branching to $ZZ$ ($\mathrm{BR} \propto g_{HZZ}^2/\Gamma_H$) gives the total Higgs width $\Gamma_H \propto \sigma(\epem \to ZH)/\mathrm{BR}(H \to ZZ)$. Similar possibilities provides the reaction $\epem \to H\nu\bar{\nu}$ where $\Gamma_H \propto \sigma (WW \to H)/\mathrm{BR}(H\to WW)$.

\subsection{Muon collider Higgs factory}
\begin{figure}[!tbp]
     \begin{center}
     \vspace*{-0.2cm}
\includegraphics[width=7.cm] {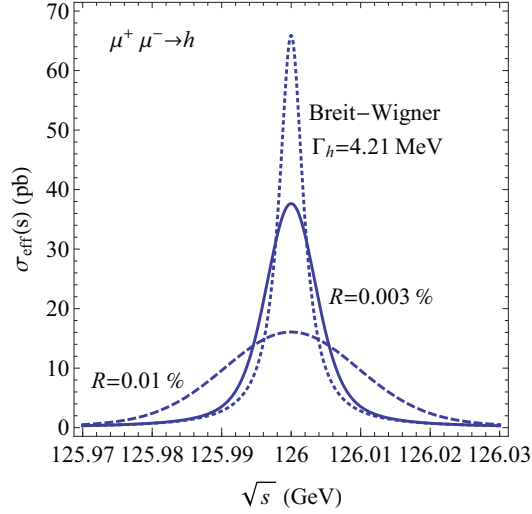}
       \vspace*{-0.3cm}
     \end{center}
     \caption{Direct measurement of the Higgs decay width at the muon collider with a high energy resolution.}
   \vspace*{0.5cm}
   \label{muon-width}
   \end{figure}

\noindent
   At muon colliders, the Higgs boson can be produced in the $s$-channel as a single resonance. The cross section of the reaction $\mu^+\mu^- \to H(126)$ is 70 pb, Fig.~\ref{muon-width}~\cite{Han}. 
In order to measure directly the Higgs width, the energy spread of muon beams should be comparable to $\Gamma_H\approx 4.2$~MeV. The relative energy spread can be reduced to $3\times 10^{-5}$ by emittance exchange at the expense of transverse emittance. The luminosity will be not high, about $10^{31} \cms$, but due to a high cross section one can produce about 5000 Higgs events per year for one IP. Scanning of the peak gives the Higgs total width with a precision of 3\%; coupling to $\mu\mu$---with accuracy of 1.5\% and the Higgs mass with a 0.1 \mevcc accuracy. All other Higgs parameters can be measured better at \epem colliders thanks to their lower physics and machine backgrounds. Studies are on-going to increase the luminosity at the Higgs up to $10^{32} \cms$.

   The muon collider Higgs factory is useful and desirable in all scenarios (with \epem colliders or without) because this technique paves the way to the highest collider energies conceivably achievable by humankind, up to 100 TeV.

\subsection{Photon colliders}
\begin{figure}[tbp]
     \begin{center}
%     \vspace*{-0.2cm}
\includegraphics[width=8cm] {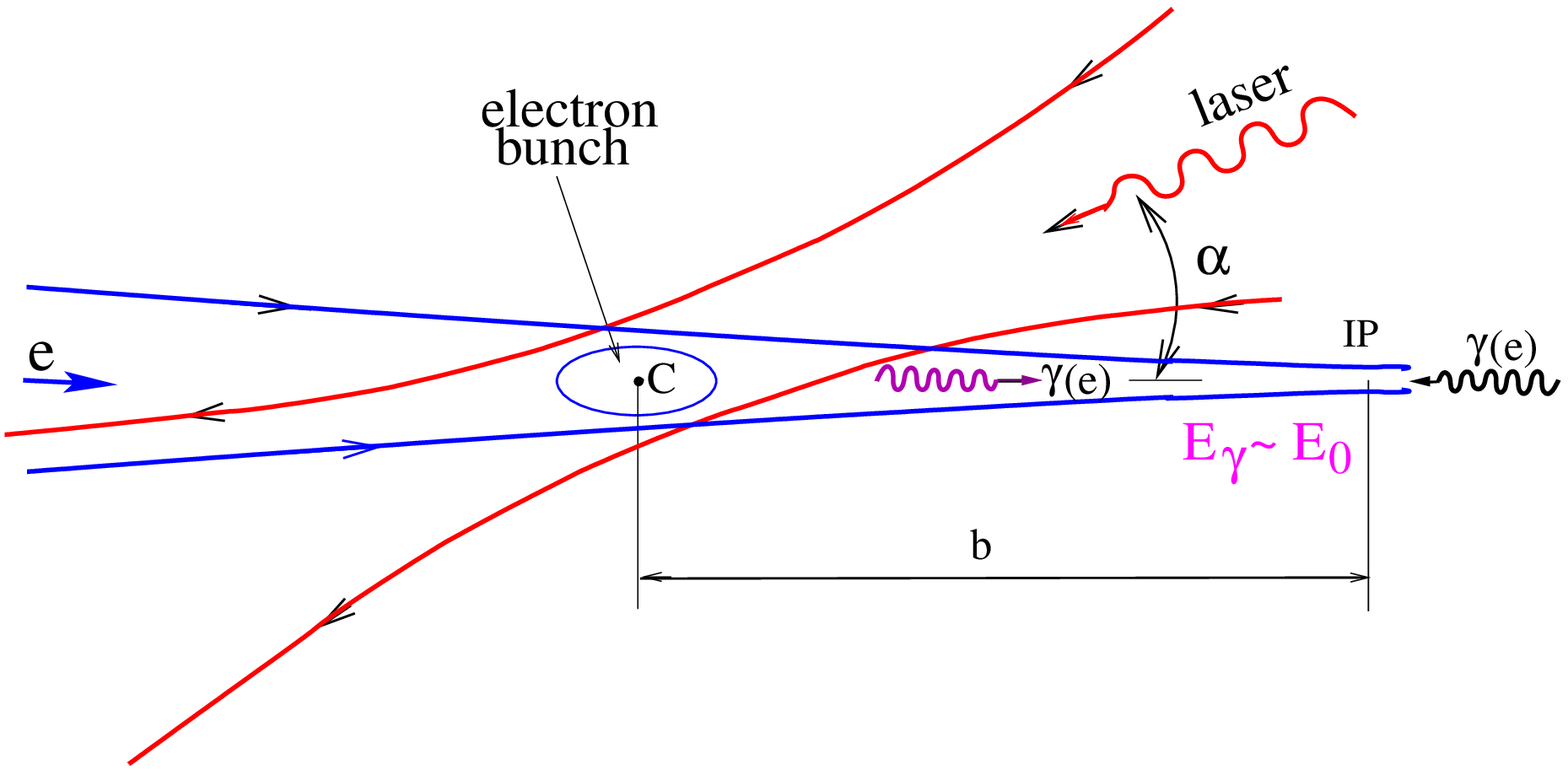} \vspace*{0cm} ~~ \includegraphics[width=6.cm] {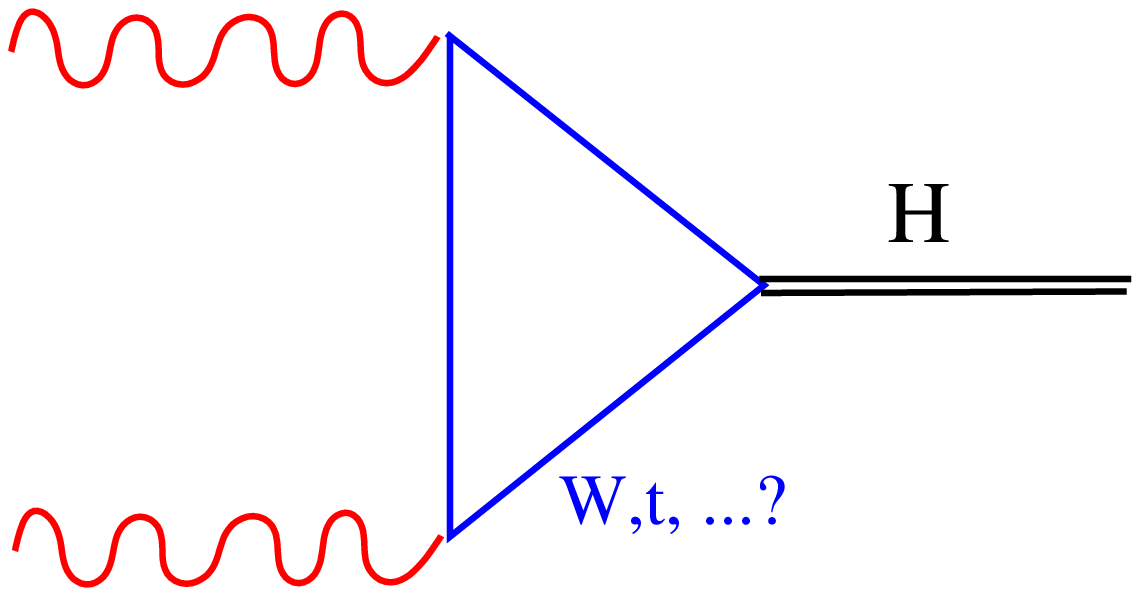}
       \vspace*{-0.3cm}
     \end{center}
     \caption{Left: scheme of \gg,\ge collider ; Right: the diagram for Higgs production in \gg collsions.}
   \vspace*{0.5cm}
   \label{plc}
   \end{figure}

\noindent
In \gg collisions, the Higgs boson is produced as a single resonance via the loop diagram (Fig.~\ref{plc}) where the leading contributions come from the heaviest charged  particles in the loop: $t$, $W$, $b$.  The measurement of this reaction's cross section can reveal the existence of yet-unknown heavy charged particles that cannot be directlty produced at colliders due to their high masses. For monochromatic photons, the cross section would be huge, about 700 pb, a factor of 10 greater than that at a muon collider. Unfortunately, at realistic photon colliders based on Compton backscattering the energy spread of the high-energy peak is about 15\% at half maximum. Even with such an energy spread, the Higgs creation cross section (for photons in this peak) is about 1 pb, a factor of 5 greater than in \epem collisions. However, the luminosity in this peak is about 10\% of the geometric $e^-e^-$ luminosity (which is approximately equal to the \epem luminosity); therefore, the effective cross section (related to $L_{ee}$) for Higgs production at photon colliders is about 200 pb, similar to that in \epem collisions.

   The Higgs boson at photon colliders can be observed in the $bb, \gg, WW$ decay modes. The Higgs \gg width $\Gamma_{\gg}$ can be measured with an accuracy of about 2\%, better than at other collider types. However, this requires the knowledge of $\mathrm{BR}(H\to b\bar{b})$, which can be measured with sufficient accuracy only at \epem colliders. Using variable photon polarizations, one can measure the Higgs boson's $CP$ properties. Although the photon collider can produce similar numbers of Higgs bosons as an \epem collider, due to the irreducible QED backgrounds one cannot detect the Higgs in the $cc, \tau\tau, \mu\mu, gg$ modes, measure directly the branchings, and see the invisible decays. Therefore, an \epem collider would be much more powerful for the Higgs study, and the photon collider will be useful only for a number of specific additional measurements---first and foremost, $\Gamma_{\gg}$, which, in fact, could be the most interesting measurement to be done at a photon collider due to its sensitivity to the possible existence of new massive charged particles.

The photon collider has always been considered a natural and fairly inexpensive addition to any LC project (because at LCs the beams are used only once). A photon collider could operate in parallel with \epem collisions (in the case of two IPs) or be implemented as the second stage of an LC project (in the case of a single IP). At the HF2012 workshop, a number of photon collider proposals for the study of the Higgs have been put forward that envision no \epem collisions at all.

   In my opinion, given the present situation with physics and funding prospects, the photon collider should be considered not as a separate Higgs factory project but as part of an \epem linear collider (ILC, CLIC, etc.).

 \subsection{Physics summary}
\noindent
   The expected accuracies with which Higgs couplings could be measured at various colliders are summarized in Table~\ref{Table1}. Comparing them with those achievable at the LHC, one can see that an \epem collider is certainly needed. The best accuracies, below 1\%, can be achieved at large ring \epem colliders, such as the TLEP. 

However, this does not mean that we should give up on energy-frontier linear colliders: new physics in the 1-3 TeV range has not yet been excluded by the LHC. Moreover, the ILC Technical Design Report is ready, and the ILC project appears to be close to approval (in Japan). It would be the right thing to do to culminate the 25 years of research and development into the superconducting LC technology (TESLA, and then ILC) by building the ILC and exploiting to the fullest its physics potential.

\begin{table*}[!tbp]
%\vspace{3mm}
\small
{\renewcommand{\arraystretch}{1.1} \setlength{\tabcolsep}{0.62mm}
\begin{tabular}{| l | c| c| c| c| c|  c|c| }  \hline
&&&&&&& \\
& LHC & HL-LHC & ILC & Full ILC & CLIC & LEP3(4IP) &  TLEP(4IP) \\
& 300 \invfb  & 3000 \invfb & 250 GeV & 250+350 & 350 GeV (500 \invfb ) & 240 GeV & 240 GeV\\
& /exp & /exp & 250 \invfb &1 TeV &1.4 TeV (1.5 \invab )& 2 \invab & 10 \invab 5yrs \\
& & & 5 yrs    & 5 yrs each  & 5 yrs each& 5 yrs & 350 GeV \\
&&&&&&& 1.4 \invab 5yrs \\ \hline
$N_H$& $1.7\times 10^7$ &$1.7\times 10^8$& $6\times10^4$ $ZH$& $10^5$ $ZH$& $7.5\times 10^4$&$4\times 10^5$ & $2\times 10^6$ $ZH$ \\
&&&&$1.4\times 10^5$ $H\nu\nu$  &$4.7\times 10^5$ $H\nu\nu$ & & $3.5\times 10^4$ $H\nu\nu$  \\ \hline
$\Delta m_H$, \mev & 100 & 50 & 35 & 35 & 100 & 26 & 7 \\ \hline
$\dG_H /\G_H $ &--&--& 10\% & 3\% & ? & 4\% & 1.3\% \\[-1mm]
& indirect & indirect &&&&& \\[-2mm]
$\dG_\inv /\G_H $ &30\%&10\%&1.5\%&1.0\%&?&0.35\%&0.15\% \\
$\dG_{\ggr} /\G_{\ggr} $ &6.5-5.1\%&5.4-1.5\%&--&5\%&?&3.4\%&1.4\% \\
$\dG_{gg} /\G_{gg} $ &11-5.7\%&7.5-2.7\%&4.5\%&2.5\%&$<$3\%&2.2\%&0.7\%\\
$\dG_{WW} /\G_{WW} $ &5.7-2.7\%&4.5-1\%&4.3\%&1\%&1\%&1.5\%&0.25\%\\
$\dG_{ZZ} /\G_{ZZ} $ &5.7-2.7\%&4.5-1\%&1.3\%&1.5\%&1\%&0.65\%&0.2\%\\
$\dG_{HH} /\G_{HH} $ &--&<30\%&--&$\sim 30$\%&$\sim 22$\%&--&--\\[-1mm]
&&&&&($\sim 11$\%, 3 TeV)&& \\[-1mm]
$\dG_\mumur /\G_\mumur $ &<30\%&<10\%&--&--&10\%&14\%&7\%\\
$\dG_\tautau /\G_\tautau $ &8.5-5.1\%&5.4-2.0\%&3.5\%&2.5\%&$\sim 3$\%&1.5\%&0.4\%\\
$\dG_{cc} /\G_{cc} $ &--&--&3.7\%&2\%&2\%&2\%&0.65\%\\
$\dG_{bb} /\G_{bb} $ &15-6.9\%&11-2.7\%&1.4\%&1\%&1\%&0.7\%&0.22\%\\
$\dG_{tt} /\G_{tt} $ &14-8.7\%&8-3.9\%&--&5\%&3\%&--&30\% \\ \hline
\end{tabular}
%\end{ruledtabular}
\vspace{-0mm}
}
\caption{Expected accuracy of Higgs-boson coupling measurements.}
\label{Table1}
\end{table*}

\section{Linear colliders}
\begin{figure}[tbhp]
     \begin{center}
     \vspace*{-0.2cm}
\includegraphics[width=14cm,clip=true] {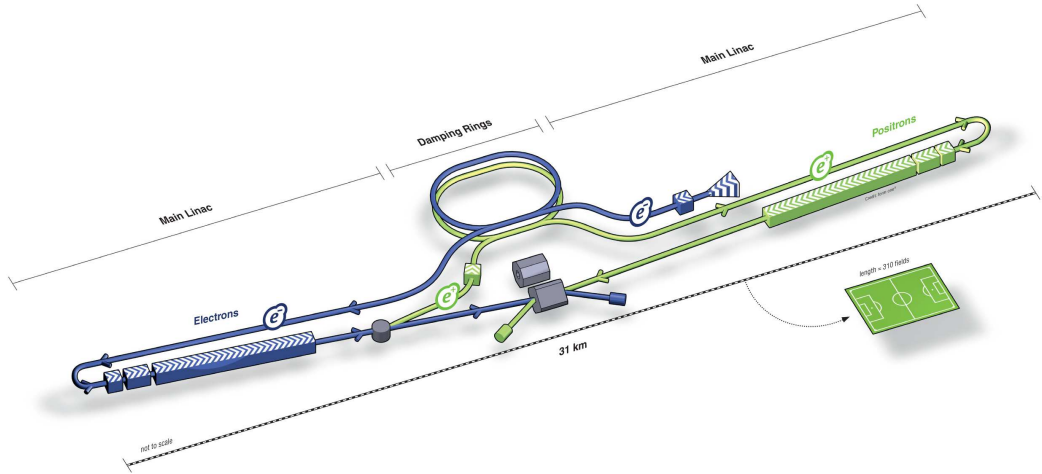}
       \vspace*{-0.7cm}
     \end{center}
     \caption{Layout of the ILC.}
   \vspace*{0.5cm}
   \label{ILC}
   \end{figure}

\begin{figure}[tbhp]
     \begin{center}
     \vspace*{-0.2cm}
\includegraphics[width=13cm] {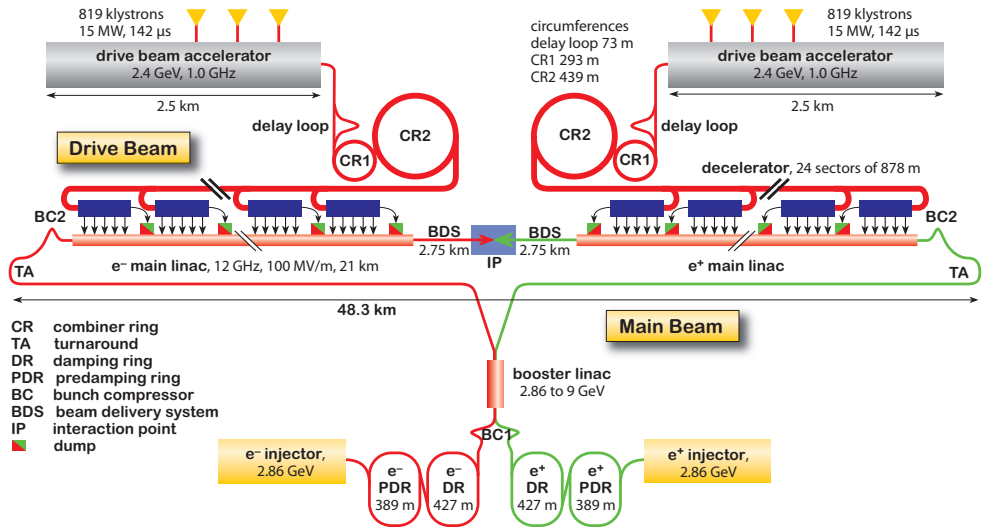}
       \vspace*{-0.5cm}
     \end{center}
     \caption{Layout of the CLIC.}
   \vspace*{0.5cm}
   \label{CLIC}
   \end{figure}

\noindent
Linear colliders have a long history. Over the years, there have been many LC projects: VLEPP, SLC, NLC, JLC, GLC, SBLC, TESLA, CLIC. Only the SLC was built, at SLAC, at the energy of the $Z$ boson. All other projects from the above list are much more ambitious: higher energy, much higher luminosity---and much higher cost, too high for any one world region to manage on its own. At present, two projects remain and are under development: the superconducting collider ILC~\cite{ILC} (based on TESLA), Fig.~\ref{ILC}, and the CLIC~\cite{CLIC}, Fig.~\ref{CLIC}, which adopts a drive-beam scheme to produce the RF for main linac. Parameters of these colliders are presented in Table~\ref{Table2}.

\begin{figure}[tbhp]
     \begin{center}
     \vspace*{-0.2cm}
\includegraphics[width=14cm] {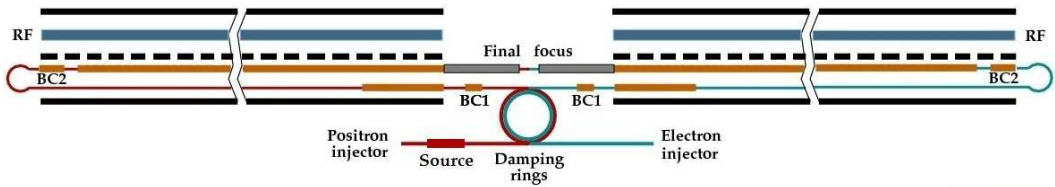}
       \vspace*{-0.5cm}
     \end{center}
     \caption{Layout of the X-band LC Higgs factory.}
   \vspace*{0.5cm}
   \label{x-band}
   \end{figure}
 \begin{table*}[!hbtp]
%\vspace{3mm}

{\renewcommand{\arraystretch}{0.9} \setlength{\tabcolsep}{2mm}
\begin{center}
\begin{tabular}{| l | l | c c c| c c  c| }  \hline
& unit && ILC &&& CLIC & \\ \hline
$2E_0$&GeV&250&500&1000&250&500&3000 \\
$L_{{\rm tot}}$&$10^{34}$\cms&0.75&1.8&4.9&1.37&2.3&5.9 \\
$L_{{\rm geom}}$&$10^{34}$\cms&0.37&0.75&2.61&0.82&1.42&4.29 \\
No. Higgs/yr($10^7$s)&1000&23&49&--&34&44&446 \\
Length&km&21&31&48&13.2&13.2&48.3 \\
$P$ (wall)&MW&128&162&301&225&272&589 \\
Pol. $e^-$/Pol. $e^+$&\%&80/30&80/30&80/30&80/0&80/0&80/0 \\
Accel. gradient&MV/m&31.5&31.5&31.5/45&40&80&100 \\
$N$ per bunch&$10^{10}$&2&2&1.74&0.34&0.68&0.372 \\
Bunches per pulse & &1312&1312&2450&842&354&312 \\
Bunch distance&ns&554&554&366&0.5&0.5&0.5 \\
Rep. rate&Hz&5&5&4&50&50&50 \\
Norm. emit. $\epsilon_{x,\,n}$&mm-mrad&10&10&10&0.66&2.4&0.66 \\
Norm. emit. $\epsilon_{y,\,n}$&mm-mrad&0.035&0.035&0.03&0.025&0.025&0.02 \\
$\beta_x$ at IP&mm&13&11&11&8&8&4 \\
$\beta_y$ at IP&mm&0.41&0.48&0.23&0.1&0.1&0.07 \\
$\sigma_x$ at IP&nm&729&474&335&150&200&40 \\
$\sigma_y$ at IP&nm&7.66&5.9&2.7&3.2&2.3&1 \\
$\sigma_z$ at IP&mm&0.3&0.3&0.225&0.072&0.072&0.044 \\
Ener. loss. $\delta E/E$&\%&0.95&4.5&10.5&1.5&7&28 \\
\hline
\end{tabular}
\end{center}
%\end{ruledtabular}
\vspace{-0mm}
}
\caption{Parameters of \epem linear colliders ILC and CLIC.}
\label{Table2}
\end{table*}

  The accelerating gradient in the ILC design is 31.5 MeV/m; at CLIC, it is up to 100 MeV/m. Correspondingly, the maximum energy achievable is 1 \tev at ILC and 3 \tev at CLIC. The total length of both the ILC and CLIC is about 48 km. The pulse structure is very different: at the ILC, the distance between bunches is about 500 ns, while at CLIC, it is a mere 0.5 ns,  which creates problems for the detector due to integration of backgrounds from multiple beam collisions.

The ILC team finished the TDR in spring of 2013; the CLIC team published the CDR in 2012. Both projects envision a 500 GeV first stage but can also start with a lower energy, $2E=250$ \gev at the ILC and 350 \gev at CLIC. For the ILC, it means that not all SC cavities would be installed. The cost of a $250 \gev$ Higgs factory is projected to be 67\% of the cost of the 500 GeV collider; 75\% if the tunnel for 500 GeV is fully constructed. At CLIC, the first stage could be based (as an alternative) on klystrons, which is somewhat cheaper. The construction of the ILC could start two years after approval (that is, potentially as early as 2016); the most probable host country is Japan. CLIC construction (if necessary) could start in or around 2022 near CERN.

In addition to ILC and CLIC, it was proposed at HF2012 (by R.~Belusevic and T.~Higo) to construct at KEK an X-band linear collider for the Higgs study with the total length of 3.6 km (Fig.\ref{x-band}). Who knows, perhaps this is indeed the best solution, despite the fact that Japan expresses a special interest in the ILC because it is based on superconducting RF technology, which has a big potential for industrial applications.

\begin{table*}[tbph]
%\vspace{3mm}
{
\renewcommand{\arraystretch}{0.9} \setlength{\tabcolsep}{2mm}
\small
\begin{center}
\begin{tabular}{| l | l | c c c| c c  c| }  \hline
& unit &LEP3&TLEP&Sup-Tristan&FNAL&IHEP&SLAC/LBNL \\
&&&&&&China& \\ \hline
$2E_0$&GeV&240&240&240&240&240&240 \\
$L_{{\rm tot}}$ per IP&$10^{34}$\cms&1&4.9&1&0.52&3.85&1 \\
Number of IPs &&2-4&2-4&1&1&1&1 \\
No. Higgs/yr($10^7$s) per IP&1000&28&130&28&14&110&28 \\
Length&km&26.7&81&40&16.2&70&26.7 \\
$P$ (wall)&MW&200&200&100&200&300&200 \\
Pol. $e^-$/Pol. $e^+$&\%&0/0&0/0&0/0&0/0&0/0&0/0 \\
$N$ per bunch&$10^{10}$&100&50&67&80&60&8 \\
Bunches per beam& &4&80&8&2&52&50 \\
$\Delta E$ per turn&GeV& 7&2.1&3.5&10.5&2.35&7 \\
$P$ per beam&MW&50&50&22.5&50&50&50 \\
Norm. emit. $\epsilon_{x,\,n}$&mm-mrad&5870&2210&90400&5321&3358&1010 \\
Norm. emit. $\epsilon_{y,\,n}$&mm-mrad&23&12&9.4&27&16.7&5 \\
$\beta_x$ at IP&mm&200&200&200&200&200&50 \\
$\beta_y$ at IP&mm&1&1&1&2&1&1 \\
$\sigma_x$ at IP&nm&71000&43000&89000&67300&53500&14700 \\
$\sigma_y$ at IP&nm&320&220&63&4.76&265&2.65 \\
$\sigma_z$ at IP&mm&3.1&1.7&1.2&2.85&1&1.5 \\
\hline
\end{tabular}
\end{center}
%\end{ruledtabular}
\vspace{-0mm}
}
\caption{Parameters of \epem circular colliders.}
\label{Table3}
\end{table*}

\section{\boldmath{Circular \epem colliders}}
\noindent
As discussed above, in late 2011 A.~Blondel and F.~Zimmermann proposed a storage-ring \epem\ Higgs factory in the LEP tunnel (LEP3). Soon after that, K.~Oide considered the Higgs factory Super-Tristan with a radius of 40-60 km around Tsukuba. He suggested the use of the crab-waist collision scheme, which, at the first glance, would give a luminosity much higher than at the ILC at 240 GeV, and still higher than the ILC even at 500 GeV. However, immediately thereafter I found that at such high energies the beam lifetime is limited by beamstrahlung~\cite{Telnov-prl}. Beamstrahlung puts an additional condition on the value of $N/(\sigma_x \sigma_z)$, and thus on the luminosity of high-energy \epem ring colliders. It turned out that the crab-waist scheme is of no benefit. Nevertheless, at the energy $2E_0=240$ \gev, as needed for the Higgs study, the luminosity could be comparable to that at linear colliders, and even higher for rings with larger radius. In addition, ring colliders can have several IPs, which further increases the total integrated luminosity. The tunnel can be used at a later time for a 100 TeV $pp$ collider. These new perspectives have triggered a great interest to circular \epem Higgs factories: the number of proposals has already reached one dozen. Parameters of some storage-ring \epem Higgs factories are presented in Table~\ref{Table3}.

\begin{figure}[!tbph]
     \begin{center}
     \vspace*{-0.2cm}
\includegraphics[width=12cm] {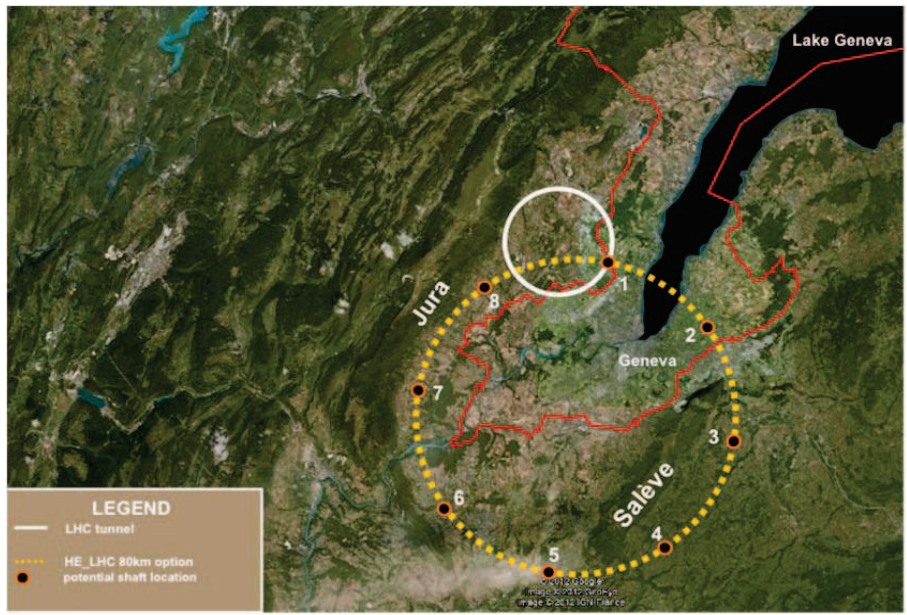}
       \vspace*{-0.3cm}
     \end{center}
     \caption{The LEP3 (white) and TLEP (yellow) rings near Geneva.}
   \vspace*{0.5cm}
   \label{TLEP}
   \end{figure}

The development of such ring colliders has just started. Some numbers in Table~\ref{Table3} need further checks and optimization. Some uncertainty is connected with the energy acceptance of the rings. The 80 km TLEP project~\cite{Blo} is supported by the CERN management; a possible location of the TLEP ring near Geneva is shown in Fig.~\ref{TLEP}. In addition to being a Higgs factory, TLEP can also serve as a $Z$ factory with a luminosity $\sim 10^{36}$ \cms. Beam polarization is not essential for the Higgs study; however, a longitudinal polariation would be highly desirable for TeraZ factory, and a transverse polarization could be used for precise energy calibration using the phenomenon of resonant depolarization of the stored beam.

\begin{figure}[!tbph]
     \begin{center}
     \vspace*{-0.2cm}
\includegraphics[width=13cm] {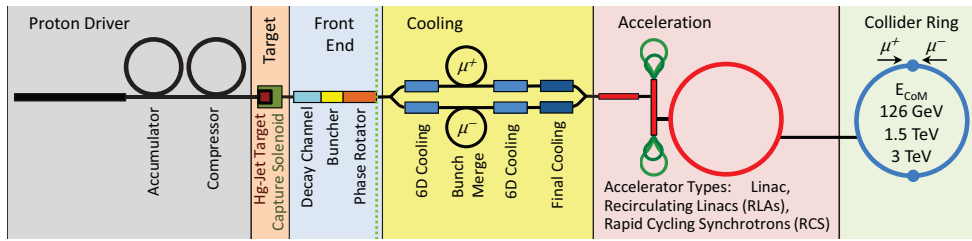}
       \vspace*{-0.3cm}
     \end{center}
     \caption{Scheme of the muon collider.}
   \vspace*{0.5cm}
   \label{muon-col}
   \end{figure}

    \begin{figure}[tbph]
     \begin{center}
     \vspace*{-0.2cm}
\includegraphics[width=9cm] {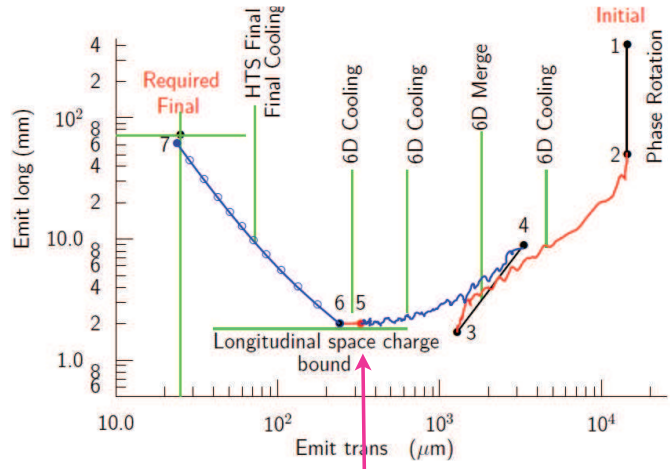}
       \vspace*{-0.3cm}
     \end{center}
     \caption{Ionization cooling procedure at muon colliders.}
   \vspace*{0.5cm}
   \label{i-cool}
   \end{figure}

\begin{table*}[!tbph]
%\vspace{3mm}
%\small
{\renewcommand{\arraystretch}{0.9} \setlength{\tabcolsep}{3mm}
\begin{center}
\begin{tabular}{| l | l | c | c| }  \hline
& unit &Low L&High L  \\ \hline
$2E_0$&GeV&126&126 \\
Luminosity per IP&$10^{34}$\cms&0.001&0.01 \\
Number of IPs &&2&2 \\
No. Higgs/yr($10^7$s) per IP&1000 &5 &50\\
Circumference&km&0.3&0.3 \\
$P$ (wall)&MW&100&125 \\
Pol. $\mu^-$ and $\mu^+$&\%&10&10-20 \\
$N$ per bunch&$10^{10}$&200&500 \\
Bunches per beam& &1&1 \\
Norm. emit. $\epsilon_{x,n}$&mm-mrad&400&200 \\
Norm. emit. $\epsilon_{y,n}$&mm-mrad&400&200 \\
$\beta_x$ at IP&mm&60&40 \\
$\beta_y$ at IP&mm&60&40 \\
$\sigma_x$ at IP&\mum&200&120 \\
$\sigma_y$ at IP&\mum&200&120 \\
$\sigma_z$ at IP&mm&60&40 \\
$\sigma_E/E$&\%&0.003&0.003 \\

\hline
\end{tabular}
\end{center}
%\end{ruledtabular}
\vspace{-0mm}
}
\caption{Parameters of the $\mu^+\mu^-$ Higgs factory.}
\label{Table4}
\end{table*}

\section{Muon colliders}
\noindent
The muon collider is attractive because the muon is a point-like particle, same as the electron but 200 times heavier; therefore, the synchrotron radiation and beamstrahlung are both negligible. The size of the ring where muons make about 2000 turns before decaying can be quite small (300~m circumference for the Higgs factory). The layout of the muon collider is shown in Fig.~\ref{muon-col}. Parameters of the $\mu^+\mu^-$ Higgs factory are presented in Table~\ref{Table4}. The luminosity is 2-3 orders of magnitude smaller than at \epem colliders, but the Higgs production cross section is about 200 times higher. The energy can be calibrated with an accuracy of about 0.1~\mev by measuring the oscillation frequency of electrons from muon decays.  The most significant challenge is making ionization cooling work. As shown in Fig.~\ref{i-cool}, 6D cooling requires a reduction by a factor of $10^{6}$. This has to be done very rapidly, before the muon decays. For the Higgs factory, the longitudinal emittance is more important for measurement of the Higgs total width, so the last stage of cooling is not needed. There are many highly challenging technical issues to be solved and demonstrated experimentally before the construction of a muon collider can start.

\section{Photon colliders}
\noindent
Photon colliders ($\gg, \gamma e$) based on one-pass linear colliders (PLCs) have been in development since 1981. A detailed description of the PLC can be found in Ref.~\cite{PLC-TESLA}. After undergoing Compton scattering at a distance $b \sim 1$~mm from the IP (Fig.~\ref{plc}), the photons have an energy close to that of the initial electrons and follow the electrons' original direction toward the IP (with a small additional energy spread of the order of $1/\gamma$). Using a modern laser with a flash energy of several joules, one can ``convert'' almost all electrons to high-energy photons. The maximum energy of the scattered photons (neglecting nonlinear effects) is
\begin{equation}
\omega_m=\frac{x}{x+1}E_0; \;\;\;\;
x \approx \frac{4E_0\omega_0}{m^2c^4}
 \simeq 15.3\left[\frac{E_0}{\tev}\right]\left[\frac{\omega_0}{\ev}\right]=
 19\left[\frac{E_0}{\tev}\right]
\left[\frac{\mum}{\lambda}\right].
\label{x}
\end{equation}

As discussed earlier in the paper, the Higgs boson is produced in \gg collisions as a single resonance.
For the Higgs factory, one can take $E_0\approx 80 \gev$ and $\lambda=1.06/3$ \mum ($x=4.3$) or
$E_0\approx 105 \gev$ and $\lambda=1.06$ \mum ($x=1.88$). The first option needs a lower electron energy but
is less preferable due to problems with the removal of too-low-energy final electrons that are deflected by
the opposing electron beam and by the detector field.

The luminosity spectrum in \gg collisions has a peak at maximum energies with $\mathrm{FWHN} \sim 15-20\%$ containing the \gg
luminosity of about 10\% of the geometric $e^-e^-$ luminosity. Typical cross sections (Higgs, charged pairs) in \gg collisions are one order
of magnitude higher than in \epem collisions, therefore the numbers of events at photon and \epem colliders are comparable.

Photon colliders are usually considered as an almost free-of-charge addition to a linear \epem collider (now, ILC and CLIC). The ``to be or not to be'' for photon colliders depends on decisions on the underlying LC projects. However, very recently a number of proposals for ``circular'' photon-collider Higgs factories without \epem collisions and without damping rings have been put forward.

\begin{figure}[!tbph]
     \begin{center}
     \vspace*{-0.2cm}
\includegraphics[width=11cm,clip=true] {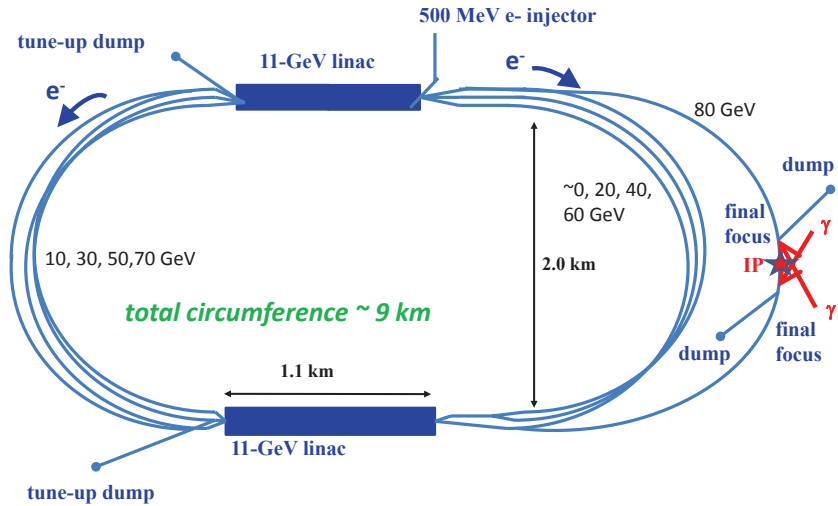}
       \vspace*{-0.3cm}
     \end{center}
     \caption{Layout of the SAPPHiRE photon collider.}
   \vspace*{0.5cm}
   \label{SAP}
   \end{figure}

{\bf SAPPHiRE}~\cite{sapphire}. 
This project is based on recirculating CW linacs (two 11 GeV linacs, 4 turns) with 80 GeV arcs, Fig.~\ref{SAP}. The chief problem in such a photon collider is the dilution of the horizontal emittance in arcs, unavailability of low-emittance polarized electron guns (though there is progress in this direction), and a problem with the removal of  electrons for the case $E_0\approx 80 \gev$ and $\lambda=1.06/3$ \mum ($x=4.3$). The collision rate is 200 kHz, 20 times higher than at ILC, which creates additional problems for the laser system.

\begin{figure}[tbph]
     \begin{center}
     \vspace*{-0.2cm}
\includegraphics[width=11cm] {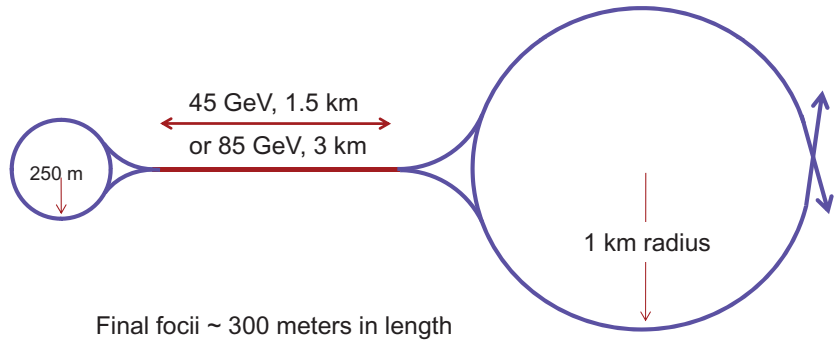}
       \vspace*{-0.3cm}
     \end{center}
     \caption{The SLC-type photon collider.}
   \vspace*{0.5cm}
   \label{SLC}
   \end{figure}

{\bf SLC-type} (T.~Rauberheimer).
This project uses 85 GeV pulsed normal-conducting or superconducting linacs (or a 45 GeV pulsed superconducting linac, twice-recirculating) to produce two 80 GeV electrons and collide them in the arcs of 1 km radius as in SLC, Fig.~\ref{SLC}.

{\bf Energy Recovery Linac-based}~\cite{Zhang}.
It uses two 50 GeV  CW  SC linacs with 50 GeV arcs. The laser flash energy (and the conversion coefficient) is much lower than usually. The luminosity is restored by a much higher beam current. It is assumed that the energy of electrons that did not interact with laser photons can be recovered by deceleration in SC linacs.

 This scheme has the following weak points: 1) beams with larger numbers of particles usuallyhave larger emittances, so the maximum luminosity will correspond to $k \sim 1$, as at normal photon colliders, 2) unscattered electrons have large loss energy due to beamstrahlung, and therefore the energy recovery of such highly non-monochromatic beams is very problematic.

\begin{figure}[!tbph]
     \begin{center}
     \vspace*{-0.2cm}
\includegraphics[width=13cm] {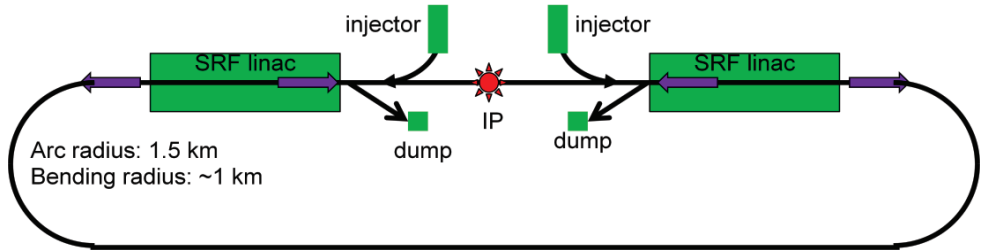}
       \vspace*{-0.3cm}
     \end{center}
     \caption{Energy-recovery linac-based photon collider.}
   \vspace*{0.5cm}
   \label{e-r-plc}
   \end{figure}

\section{Laser systems for photon colliders}
\begin{figure}[!tbph]
     \begin{center}
%     \vspace*{-0.2cm}
\includegraphics[width=10cm] {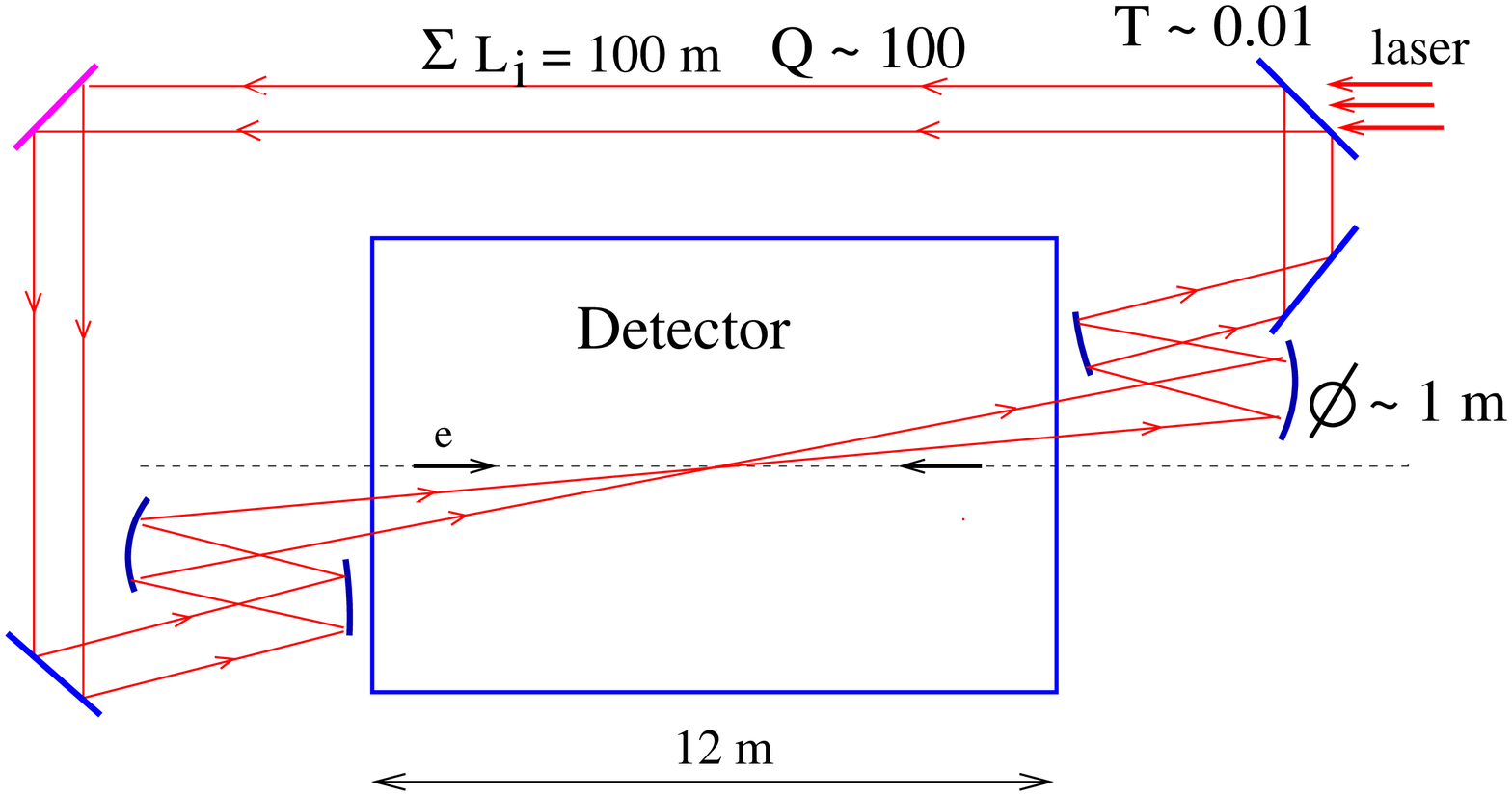}
       \vspace*{-0.5cm}
     \end{center}
     \caption{Optical cavity laser system for the photon collider at ILC.}
   \vspace*{0.5cm}
   \label{cavity}
   \end{figure}

\noindent
 The requirements on the laser system for the PLC are as follows: flash energy about 10 J, duration $\sim$ 1 ps, the wavelength $\lambda \sim 1$ \mum, and the pulse structure similar to that for the electron beams. In the case of single use of laser pulses, the average energy of each of the two lasers should be of the order of 100 kW, both for ILC and CLIC. At the ILC, the distance between the bunches is large, about 100-150 m, which makes possible the use of an external optical cavity (Fig.~\ref{cavity}), which, in turn, can reduce the required laser power by a factor of $Q \sim 100$. At CLIC, the distance between the bunches is only 15 cm, and so having an optical cavity is not possible---one needs a very powerful one-pass laser.

 \begin{figure}[htp]
     \begin{center}
     \vspace*{-0.2cm}
\includegraphics[width=7.3cm] {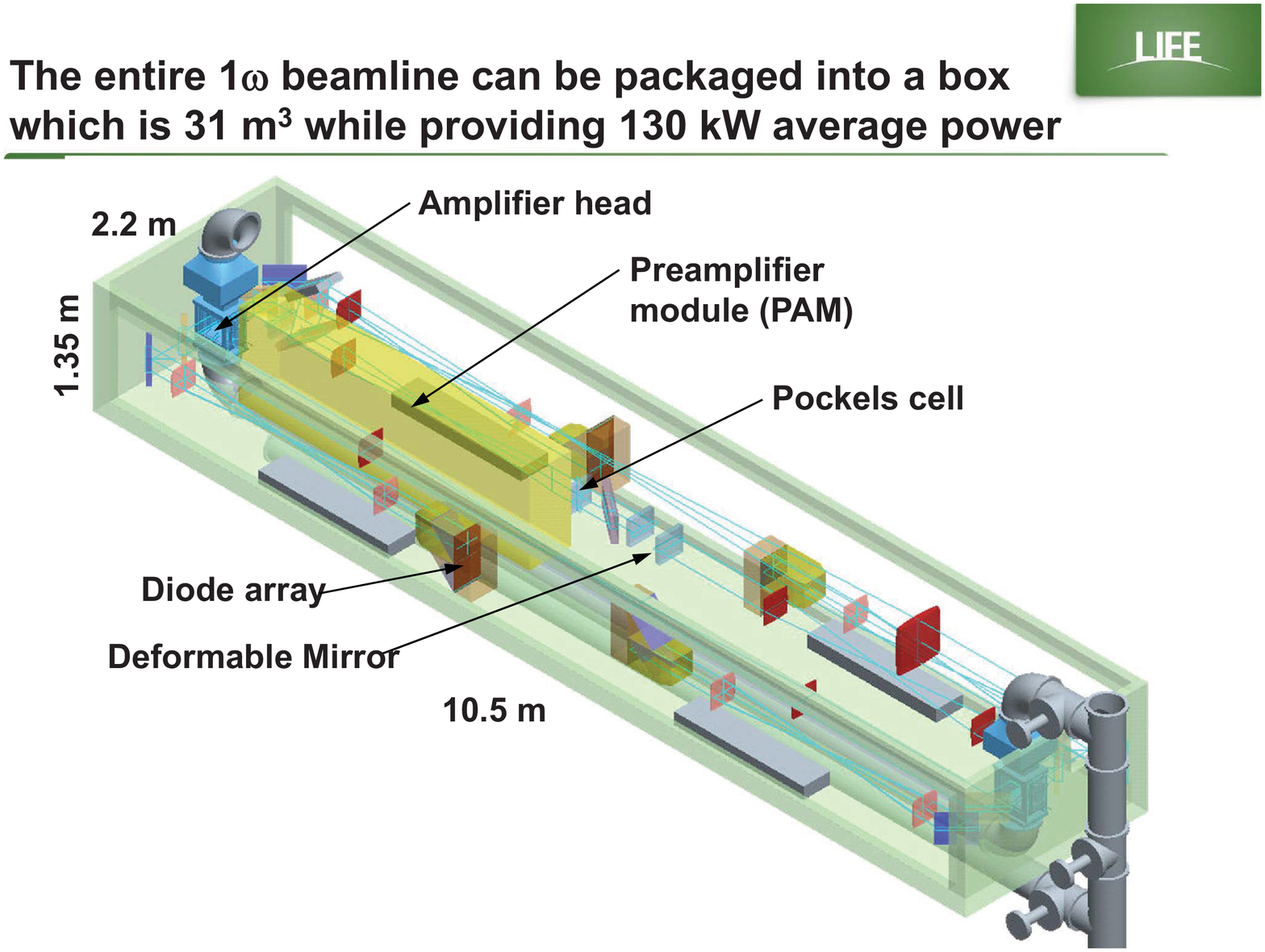} ~~ \includegraphics[width=7.3cm] {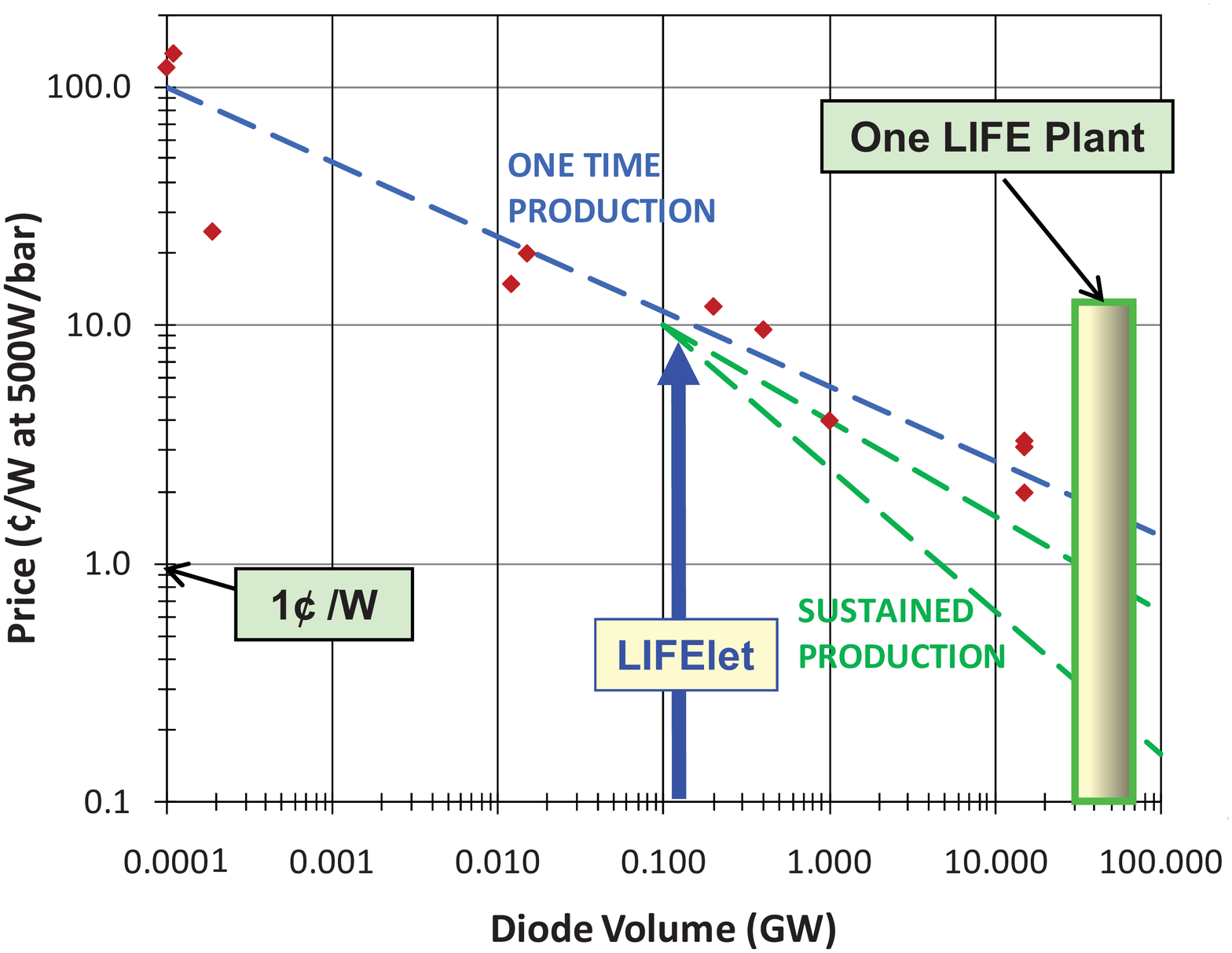}
       \vspace*{-0.3cm}
     \end{center}
     \caption{Left: One laser for the inertial fusion project LIFE at LLNL, suitable for photon colliders at both ILC and CLIC. Right: industrial price of diodes per one watt.}
   \vspace*{0.5cm}
   \label{LLNL}
   \end{figure}

Good news on lasers for the photon collider came from LLNL~\cite{Bayramian}. LLNL has developed a highly advanced laser system for the inertial fusion project LIFE. It consists of 384 lasers, each of which produces 8 kJ pulses at the rate of 16 Hz. The average power of one laser is 130 kW, similar to that required for the photon collider. It is necessary only to split 8 kJ pulses to many 10 J pulses, which is not a big problem. In addition, these pulses should be compressed to 1 ps duration using the pulse chirped techniques. Fig.~\ref{LLNL} (left) shows one of LIFE lasers. Its volume is $31 \mathrm{m}^3$. The progress in such lasers is connected with the reduction of the price of pumping diodes. The diode cost for one laser at the mass diode production for the first electrical plants will be \$2.3 million. If so, one can use such a laser both for ILC and CLIC without any enhancement in optical cavities.

    \section{Conclusion}
\noindent
    The discovery of the Higgs boson has given momentum to several new high-energy accelerator projects, has triggered active discussions of a number of Higgs factory designs. The ILC is very close to approval. A large circular \epem collider also looks very attractive due to its higher luminosity and further use of the tunnel for the next (last?) $pp$ collider. The muon collider is also very promising but needs many more years of R\&D to prove its feasibility. The CLIC project could also enter the game if the LHC finds new physics in 1-3 TeV region.

\section*{Acknowledgement}
\noindent
I would like to thank A.~Blondel, W.~Chao,    E.~Levichev, M.~Koratzinos, S.~Nagaitsev, K.~Oide, V.~Shiltsev, A.~Skrinsky, K.~Yokoya,  M.~Zanetti,  F.~Zimmermann.

 This work was supported by Russian Ministry of Education and Science.

\end{document}